\documentclass{webofc}

\usepackage[varg]{txfonts} 
\usepackage{hyperref}
\usepackage{url}
\hypersetup{colorlinks=true,citecolor=blue,urlcolor=blue,linkcolor=blue}
\usepackage[T1]{fontenc}
\usepackage{tikz-cd}
\let\a=\alpha   \let\b=\beta      

\begin{document}

\title{
 The Gravitational Form Factor of the Pion and Proton\\ and the Conformal Anomaly  \footnote{Presented at QCD@Work by Dario Melle and Riccardo Tommasi, 18 - 21 June 2024, Trani, Italy\\} }

\author{\firstname{Claudio} \lastname{Corianò}\inst{1,2}\fnsep\thanks{\email{claudio.coriano@le.infn.it }} \and
        \firstname{Stefano} \lastname{Lionetti}\inst{1,2}\fnsep\thanks{\email{stefano.lionetti@unisalento.it}}
\and
 \firstname{Dario} \lastname{Melle}\inst{1,2}\fnsep\thanks{\email{dario.melle@le.infn.it}}
\and
 \firstname{Riccardo} \lastname{Tommasi}\inst{1,2}\fnsep\thanks{\email{riccardo.tommasi@unisalento.it}}
}

\institute{Dipartimento di Matematica e Fisica, Universit\`{a} del Salento 
	and INFN Sezione di Lecce, Via Arnesano 73100 Lecce, Italy
\and
          National Center for HPC, Big Data and Quantum Computing, Via Magnanelli 2, 40033 Casalecchio di Reno, Italy
          }

\abstract{We analyze the hard scattering amplitude of gravitational form factors (GFFs) of hadrons within QCD factorization at large momentum transfers, focusing on their conformal field theory (CFT) description. These form factors are key to studying quark and gluon angular momentum in hadrons, connected to Mellin moments of Deeply Virtual Compton Scattering (DVCS). The analysis uses diffeomorphism invariance and conformal symmetry in momentum space. A non-Abelian $TJJ$ 3-point function at $O(\alpha_s^2)$ reveals a dilaton interaction in the $t$-channel. We present a parameterization relevant for future DVCS/GFF experiments at the Electron-Ion Collider (EIC).

}
\maketitle

\section{Introduction}
\label{intro}
The experimental study of the gravitational form factors (GFFs) of the proton and pion provides valuable nonperturbative insights into their interaction with the QCD energy-momentum tensor (EMT), offering a window into the internal distributions of energy, spin, pressure, and shear forces. GFFs are formulated through the matrix elements of the EMT between hadronic states, allowing for a detailed characterization of the hadron's internal structure. Quantum corrections induce a breaking of conformal symmetry due to the trace anomaly of the stress-energy tensor 

\begin{equation}
\label{anom1}
    	\hat{T}_{\!\mu}^{\,\mu} \equiv 
	{\beta(g)}\;F^{a,\mu\nu}{F^{a,}}_{\!\!\mu\nu}
    	+(1+\gamma_m)\sum_qm_q\bar\psi_q\psi_q \;,
	\end{equation}
where $\beta(g)$ is the $\beta$-function of QCD and $\gamma_m$ is 
the anomalous dimension of the mass operator. The matrix elements of the EMT for a spin 1/2 hadron with momentum \(P\) can be expressed in terms of the GFFs as

\begin{figure}[t]
\label{ff1}
\begin{center}
\rotatebox{-90}{\includegraphics[scale=0.4]{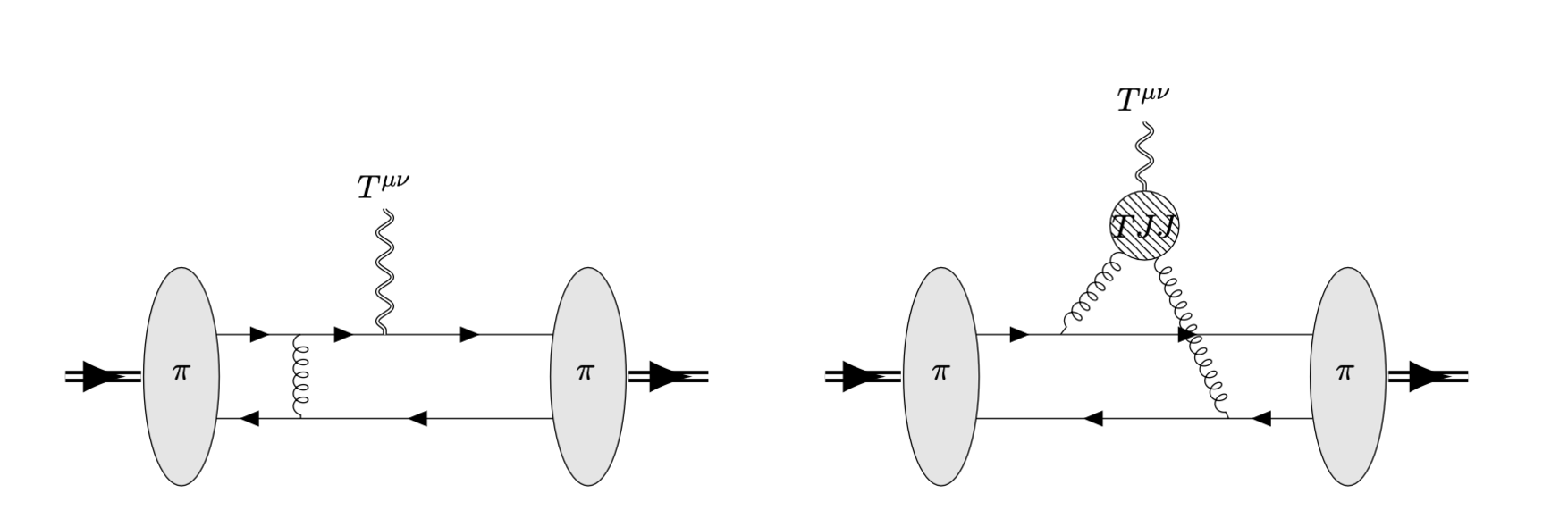}}
\caption{Factorization picture of the GFF of the pion. Leading order (left). The contribution of the conformal anomaly revealed by the $TJJ$ vertex (right).}
\end{center}
\end{figure}
\begin{figure}[t]
\begin{center}
\label{dvcs}
\rotatebox{0}{\includegraphics[scale=0.9]{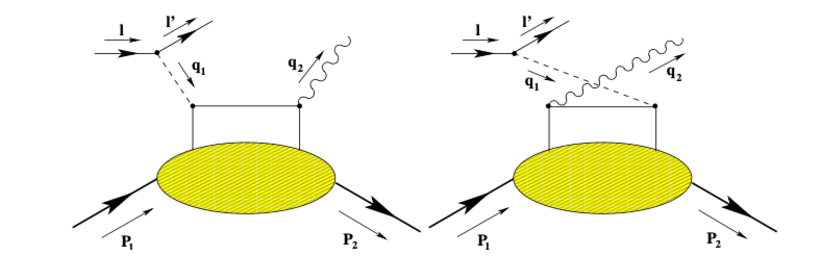}}
\caption{Factorization in the DVCS process.}
\end{center}
\end{figure}


\begin{equation}
\label{prime}
\langle p^\prime,s^\prime| T_{\mu\nu}(0) |p,s\rangle
    = \bar{u}^\prime \left[
      A(t)\,\frac{\gamma_{\{\mu} P_{\nu\}}}{2}
    + B(t)\,\frac{i\,P_{\{\mu}\sigma_{\nu\}\rho}\Delta^\rho}{4 M}
    + D(t)\,\frac{\Delta_\mu\Delta_\nu-g_{\mu\nu}\Delta^2}{4M}
    + {M}\,\sum_{\hat{a}} {\bar{c}}^{\hat{a}}(t)\,g_{\mu\nu} \right]u
\end{equation}

where \( u(p) \) and \( \bar{u}(p') \) are proton spinors, \( P \) is the average momentum, and \( \Delta \) is the momentum transfer. At large momentum transfers, the amplitude decreases rapidly in accordance with quark and gluon counting rules. 
The general idea behind quark-gluon counting rules is to count the minimum number of constituents (quarks and gluons) participating in the hard scattering process, and then determine how the amplitude of the process scales with the momentum transfer, at large values of momentum transfer $t=q^2$, where 
$q$ is the momentum of the graviton. Nevertheless, it remains accessible within the factorization framework using pion and proton distribution amplitudes, as this approach can be adapted from studies of electromagnetic form factors \cite{Li:1992nu}. 
One of the major advantages of the EIC is its exceptionally high luminosity, which is expected to reach up to $10^{34} cm^2 s^{-1}$. This high luminosity compensates for the rapid decrease in cross-sections typical of exclusive processes, allowing for a significant number of exclusive events to be detected even at large momentum transfers, where the cross-sections are very small. The large number of collisions ensures that even rare events are observable.\\
 We investigate the hard scattering behavior of gravitational form factors (GFFs) at large momentum transfers and higher perturbative orders, where the factorization framework for exclusive processes applies. As shown in Fig. \ref{ff1}, this method uncovers a conformal anomaly at the TJJ (graviton/gluon-gluon) vertex at one loop, marked by the emergence of a dilaton interaction mediated by the anomaly. Our goal is to indicate an optimal parameterization of the hard scattering amplitude identifying the key form factors from which the dilaton interaction, though subleading in the pertubative expansion, can be singled out at future experiments. Additional details and references of our analysis can be found in \cite{Coriano:2024qbr}.\\
 At first sight, a key challenge, though, is the absence of a graviton to directly probe such matrix elements. 
Gravitational form factors (GFFs) are, however, closely linked to deeply virtual Compton scattering (DVCS) processes, as described in \cite{Ji:1996nm}. In DVCS (see Fig. \ref{dvcs}), an electron ($e$) scatters off a nucleon ($N$) in the reaction $eN \to e' N' \gamma$, producing a photon in the final state, controlled by nonforward parton distributions  \cite{Ji:1996nm} \cite{Radyushkin:1996ru}. This process draws significant interest for both pions and protons. In DVCS, a high-energy electron interacts with a hadron, such as a proton or pion, via the exchange of a virtual photon, which then radiates a real photon. By measuring the DVCS invariant amplitudes it is possible to extract informations on the form factors of \eqref{prime}. 
We examine this perturbative correction extending the formalism of conformal field theory in momentum space (CFT$_p$). We propose a decomposition of the hard scattering amplitude that enables the extraction of the dilaton contribution to these matrix elements. The dilaton can be interpreted as the 
Nambu Goldstone mode of broken conformal invariance, whose effective interaction can be summarised at phenomenological level in the local form ($\Lambda^{-1}(\varphi FF)$), with $\Lambda$ a typical hadronic scale. Our analysis is built around the exact expression of this interaction in perturbative QCD, which is nonlocal, and can be investigated using the conformal anomaly effective action.
We apply methods developed for the analysis of such effective action in gravity \cite{Coriano:2020ees}, specifically adapted to QCD.

\section{The CFT-induced parameterization of the pion and proton form factors}
Dilaton-like $t$-channel exchanges in the $TJJ$ correlator were initially studied perturbatively in QED \cite{Giannotti:2008cv, Armillis:2009pq} and QCD \cite{Armillis:2010qk} to explore the coupling of gravity to Standard Model fields \cite{Coriano:2011zk}. In this work, we focus on the off-shell expansion of the non-Abelian $TJJ$ correlator, previously examined in momentum space through perturbative QCD for on-shell gluons, as detailed in \cite{Armillis:2010qk}. For a more recent discussion of the trace anomaly in QCD, we refer to \cite{Hatta:2018sqd}. In the CFT framework, $J$ denotes a non-Abelian vector current in four dimensions ($d=4$), while $T$ represents the gauge-fixed stress-energy tensor in QCD. To integrate this interaction into the generalized form factors (GFFs) of hadrons, we extend the on-shell analysis from \cite{Armillis:2010qk}, employing the CFT$_p$ formalism.
We consider the decomposition of the operators $T$ and $J$ in terms of their transverse traceless part and longitudinal (local) one, separating the quark and gluon parts, that, as we are going to see, behave differently under the application of the conformal constraints. We define, for generic momenta $p_i$, ($i=1,2,3)$
	\begin{align}
		T^{\mu_i\nu_i}(p_i)&\equiv t^{\mu_i\nu_i}(p_i)+t_{loc}^{\mu_i\nu_i}(p_i),\label{decT}\\
		J^{a_i \, \mu_i}(p_i)&\equiv j^{a_i \, \mu_i}(p_i)+j_{loc}^{a_i \, \mu_i}(p_i),\label{decJ}
	\end{align}
	where
	
	\begin{align}
		\label{loct}
		t^{\mu_i\nu_i}(p_i)&=\Pi^{\mu_i\nu_i}_{\alpha_i\beta_i}(p_i)\,T^{\alpha_i \beta_i}(p_i), &&t_{loc}^{\mu_i\nu_i}(p_i)=\Sigma^{\mu_i\nu_i}_{\alpha_i\beta_i}(p)\,T^{\alpha_i \beta_i}(p_i),\\
		j^{a_i \, \mu_i}(p_i)&=\pi^{\mu_i}_{\alpha_i}(p_i)\,J^{a_i \, \alpha_i }(p_i), &&\hspace{1ex}j_{loc}^{a_i \, \mu_i}(p_i)=\frac{p_i^{\mu_i}\,p_{i\,\alpha_i}}{p_i^2}\,J^{a_i \, \alpha_i}(p_i).
	\end{align}
having introduced the transverse-traceless ($\Pi$), transverse $(\pi)$, longitudinal ($\Sigma$) projectors. Their explicit expressions can be found in \cite{Coriano:2024qbr}.\\
We adopt the following convention for the momenta: the momenta of the two gluons are denoted by \( p_1 \) and \( p_2 \), while the momentum of the graviton coupling to the energy-momentum tensor \( T \) is denoted by \( q \). Additionally, we define the tensor \(\hat{\pi}^{\mu\nu} \equiv q^2 g^{\mu\nu} - q^\mu q^\nu\), and introduce the notation \(\pi^{\mu\nu} \equiv ({1}/{q^2})\, \hat{\pi}^{\mu\nu}\). We can enumerate all possible tensor that can appear in the transverse traceless part preserving the symmetry of the 
	correlator 
	\begin{align}
		\langle t^{\mu \nu }(p_1)j^{a \, \alpha}(p_2)j^{b \, \beta}(p_3)\rangle_q & =
		{\Pi_1}^{\mu \nu}_{\mu_1 \nu_1}{\pi_2}^{\alpha}_{\alpha_1}{\pi_3}^{\beta}_{\beta_1}
		\left( A_{1}^{(q) a b} \ p_2^{\mu_1 }p_2^{\nu_1}p_3^{\alpha_1}p_1^{\beta_1} + 
		A_{2}^{(q) a b}\ \delta^{\alpha_1 \beta_1} p_2^{\mu_1}p_2^{\nu_1} + 
		A_{3}^{(q) a b}\ \delta^{\mu_1\alpha_1}p_2^{\nu_1}p_1^{\beta_1}\right. \notag\\
		& \left. + 
		A_{3}^{(q) a b}(p_2\leftrightarrow p_3)\delta^{\mu_1\beta_1}p_2^{\nu_1}p_3^{\alpha_1}
		+ A_{4}^{(q) a b}\  \delta^{\mu_1\beta_1}\delta^{\alpha_1\nu_1}\right)\label{DecompTJJ}
	\end{align}
and decompose the $TJJ$ contribution in the form
\begin{equation}
\begin{aligned}
	& \langle T^{\mu \nu}(q)  J^{ a\alpha}(p_1) J^{ b\beta}(p_2)\rangle=\langle t^{\mu \nu}(q)  j^{ a\alpha}(p_1) j^{ b\beta}(p_2)\rangle+\langle t^{\mu \nu }(q)j_{loc}^{a  \a}(p_1)j^{b  \b}(p_2)\rangle_g +\langle t^{\mu \nu }(q)j^{a  \a}(p_1)j_{loc}^{b  \b}(p_2)\rangle_g\\
	& \qquad+2 \mathcal{I}^{\mu \nu \rho }(q)\left[\delta_{[\rho}^{\beta} p_{2 \sigma ]}\langle J^{ a\alpha}({p}_1) J^{ b\sigma}(-{p}_1)\rangle+\delta_{[\rho}^{\alpha} p_{1 \sigma]}\langle J^{ b\beta}({p}_2) J^{ a\sigma}(-p_2)\rangle\right] 
	+\frac{1}{ 3 \, q^2} \hat\pi^{\mu \nu}(q) \left[\mathcal{A}^{\alpha \beta a b}+\mathcal{B}^{\alpha \beta a b}_g\right],
\end{aligned}
\label{res}
\end{equation}
where the traceless sector is
\begin{equation}
\begin{aligned}
 \langle T^{\mu \nu}(q)  J^{ a\alpha}(p_1) J^{ b\beta}(p_2)\rangle_{tls}=&
\langle t^{\mu \nu}(q)  j^{ a\alpha}(p_1) j^{ b\beta}(p_2)\rangle+\langle t^{\mu \nu }(q)j_{loc}^{a  \a}(p_1)j^{b  \b}(p_2)\rangle_g +\langle t^{\mu \nu }(q)j^{a  \a}(p_1)j_{loc}^{b  \b}(p_2)\rangle_g\\
	& \qquad+2 \mathcal{I}^{\mu \nu \rho }(q)\left[\delta_{[\rho}^{\beta} p_{2 \sigma ]}\langle J^{ a\alpha}({p}_1) J^{ b\sigma}(-{p}_1)\rangle+\delta_{[\rho}^{\alpha} p_{1 \sigma]}\langle J^{ b\beta}({p}_2) J^{ a\sigma}(-p_2)\rangle\right], 
\end{aligned}
\label{ali}
\end{equation}
and the trace part is 

\begin{equation}
\label{trc}
\langle T^{\mu \nu}(q)  J^{ a\alpha}(p_1) J^{ b\beta}(p_2)\rangle_{tr}=\frac{1}{ 3 \, q^2} \hat\pi^{\mu \nu}(q) \left[\mathcal{A}^{\alpha \beta a b}+\mathcal{B}^{\alpha \beta a b}_g\right].
\end{equation}
The trace part contains the anomaly contribution 
$\mathcal{A}^{\alpha \beta a b}=\mathcal{A}^{\alpha \beta}\delta^{ab}$ ( $a$ and $b$ are color indices) and a second term proportional to the gluon equations of motion $\mathcal{B}^{\alpha \beta a b}_g$. This is not to be considered part of the trace anomaly although it is part of the trace of the correlator since the trace operation with a Minkowski metric $\eta_{\mu\nu}$ gives
\begin{equation}
\eta_{\mu\nu}\langle T^{\mu \nu}(q)  J^{ a\alpha}(p_1) J^{ b\beta}(p_2)\rangle_{tr}= \left[\mathcal{A}^{\alpha \beta a b}+\mathcal{B}^{\alpha \beta a b}_g\right].
\end{equation}
The first tensor term $\mathcal{A}$ defines the anomaly contribution, as given by the first term on the right-hand-side of \eqref{anom1}
\begin{equation}
\mathcal{A}^{\alpha\beta a b}(p_1,p_2)= A_n \delta^{a b} u^{\alpha\beta}(p_1,p_2),
\end{equation}
and explicitly as 
\begin{equation}
\mathcal{A}^{\alpha\beta ab} = {1 \over 3} \, \,  \, \, \frac{g_s^2}{16\pi^2} \, (11C_A - 2 n_f) \delta^{ab}u^{\alpha \beta}(p_1,p_2),
\end{equation}
which is proportional to the QCD $\beta$ function for $n_f$ massless flavours. 
We have defined the tensor structure
\begin{equation}
u^{\a \b} (p_1,p_2)\equiv (p_1\cdot p_2) g^{\alpha\beta} - p_2^\alpha p_1^\beta,
\end{equation}
that summarizes the conformal anomaly contribution, being the Fourier transform of the anomaly term at $O(g^2)$

\begin{equation}
u^{\alpha\beta}(p_1,p_2) = -\frac{1}{4}\int\,d^4x\,\int\,d^4y\ e^{ip_1\cdot x + i p_2\cdot y}\ 
\frac{\delta^2 \{F^a_{\mu\nu}F^{a\mu\nu}(0)\}} {\delta A_{\alpha}(x) A_{\beta}(y)}\vline_{A=0} \,.
\label{locvar}\\
\end{equation}
The anomaly form factor characterising the exchange of a 
dilaton pole in the $TJJ$ is then distilled from \eqref{trc} in the form 

\begin{equation}
  \frac{\beta}{q^2} \delta^{ab} \subset \langle T^{\mu \nu}(q)  J^{ a\alpha}(p_1) J^{ b\beta}(p_2)\rangle,
\end{equation}
where the $1/q^2$ anomaly pole, interpreted as the signature of a dilaton exhange in perturbative QCD, has been extracted from the longitudinal projector $\pi^{\mu\nu}$ of the last term on the right-hand side of \eqref{res}.
The analysis of the perturbative amplitude is constrained by ordinary Ward identities, Slavnov-Taylor identities (STIs) and anomalous conformal Ward identities (CWIs) that we discuss in our work. Eq. \eqref{res}, which is one of the results of our analysis, shows that the structure of the effective vertex corresponding to the $TJJ$ 
correlator is modified compared to the ordinary CFT case, 
with modifications affecting both the longitudinal and the trace sectors. In the non-Abelian case, the gauge-fixing and ghost sectors of QCD break conformal symmetry. Consequently, the reconstruction of the correlator using CFT$_p$ permits the inclusion of non-conformal terms. These terms do not appear in the decomposition of non-Abelian correlators, as Slavnov-Taylor identities (STIs) are not required in the  conformal framework. This distinction arises because the general CFT analysis is not tied to a specific Lagrangian, unlike in perturbative QCD. \\
The pole for off-shell gluons can be extracted from the trace sector by following the steps we have outlined. For the quark contributions, we utilize CFT$_p$, which is constrained solely by the standard Ward identities (WIs) and conformal Ward identities (CWIs). These serve as a guiding framework for the full expansion of the correlator. Additionally, we account for the extra contributions arising from the exchange of virtual gluons and ghosts within the loops of the $TJJ$ vertex.
One may define the quantum averaged stress energy tensor, with just the pole-term included due to the $1/q^2$ exchange 
 \begin{equation}
T^{\mu\nu}_{anom}(z) =
\beta(g) \left(g^{\mu\nu}\Box - \partial^{\mu}\partial^{\nu}\right)_z \int\,d^4x'\, \Box_{z,x'}^{-1}
\left[F^a_{a\alpha\beta}F^{a \alpha\beta}\right]_{x'}\,,
\label{Tanom}
\end{equation}
from which one may extract the anomaly-induced vertex at trilinear level in the specific form
 
 \begin{eqnarray}
 \Gamma_{\textrm{anom}}^{\mu\nu\alpha\beta\, a b}(p_1,p_2) &=& \int\,d^4x\,\int\,d^4y\, e^{ip_1\cdot x + i p_2\cdot y}\,\frac{\delta^2 T^{\mu\nu}_{anom}(0)}{\delta A^a_{\alpha}(x) A^b_{\beta}(y)}  \nonumber \\
&=& \beta(g)\frac{1}{3\, q^2} \left( g^{\mu\nu}q^2 - q^{\mu}q^{\nu}\right) u^{\alpha\beta}(p_1,p_2)\delta^{a b}
\label{Gamanom}
\end{eqnarray}
 with a trace anomaly 
\begin{equation}
g_{\mu\nu}\Gamma^{\mu\nu\alpha\beta a b }(p,q)\big\vert_{A=0} =\beta(g)
\,u^{\alpha\beta}(p_1,p_2)\delta^{a b}.\,
\label{Gamanomtr}
\end{equation}
By differentiating $T^{\mu\nu}_{\text{anom}}(0)$ to higher orders with respect to the external classical gluon field, we derive analogous relations for the 4-point function ($TJJJ$) appearing in the gauge covariant expansion of the anomaly contribution \cite{Coriano:2024qbr}. 

\begin{figure}[t]
\begin{center}
\rotatebox{-90}{\includegraphics[scale=0.1]{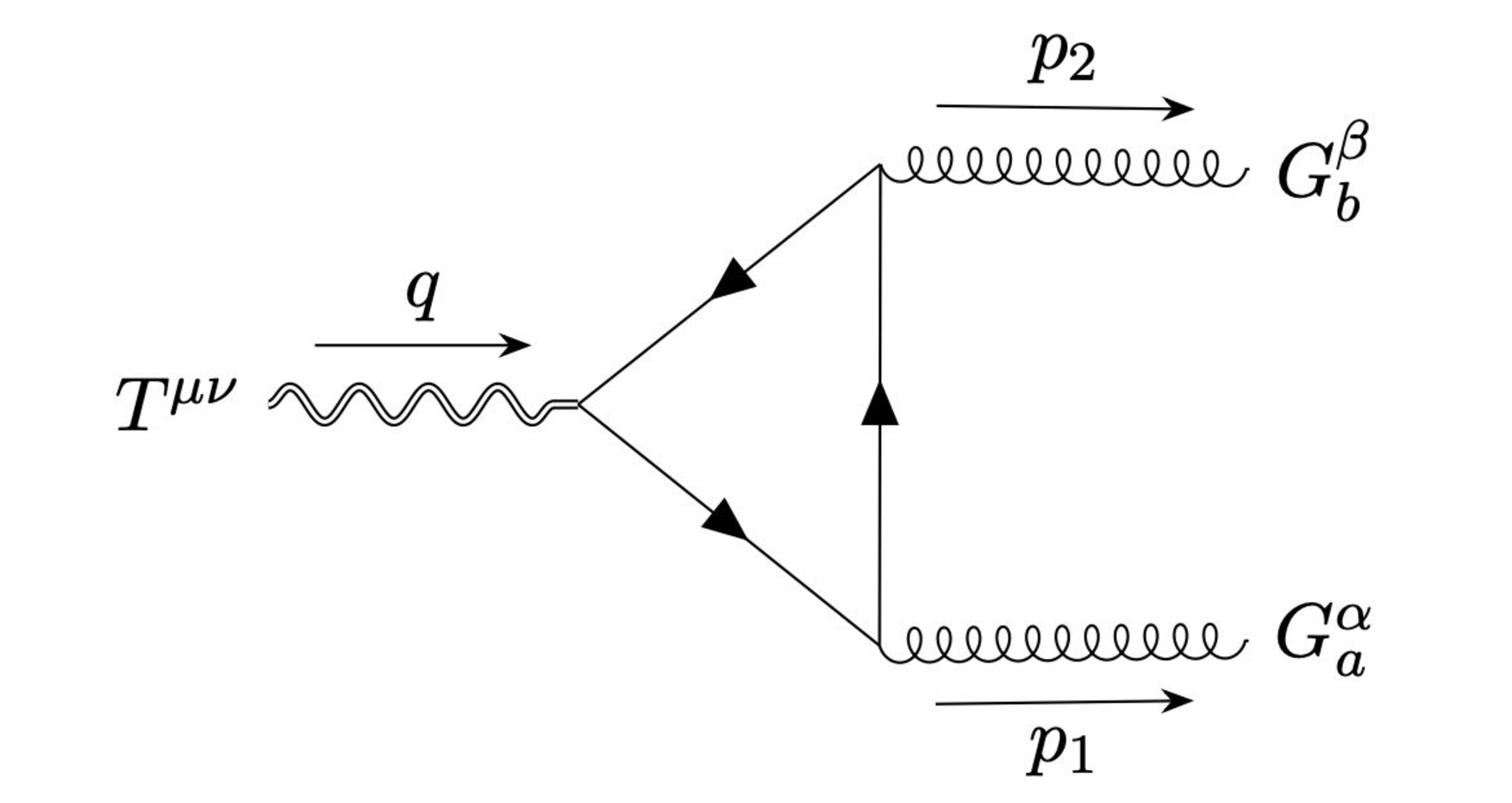}}
\rotatebox{-90}{\includegraphics[scale=0.1]{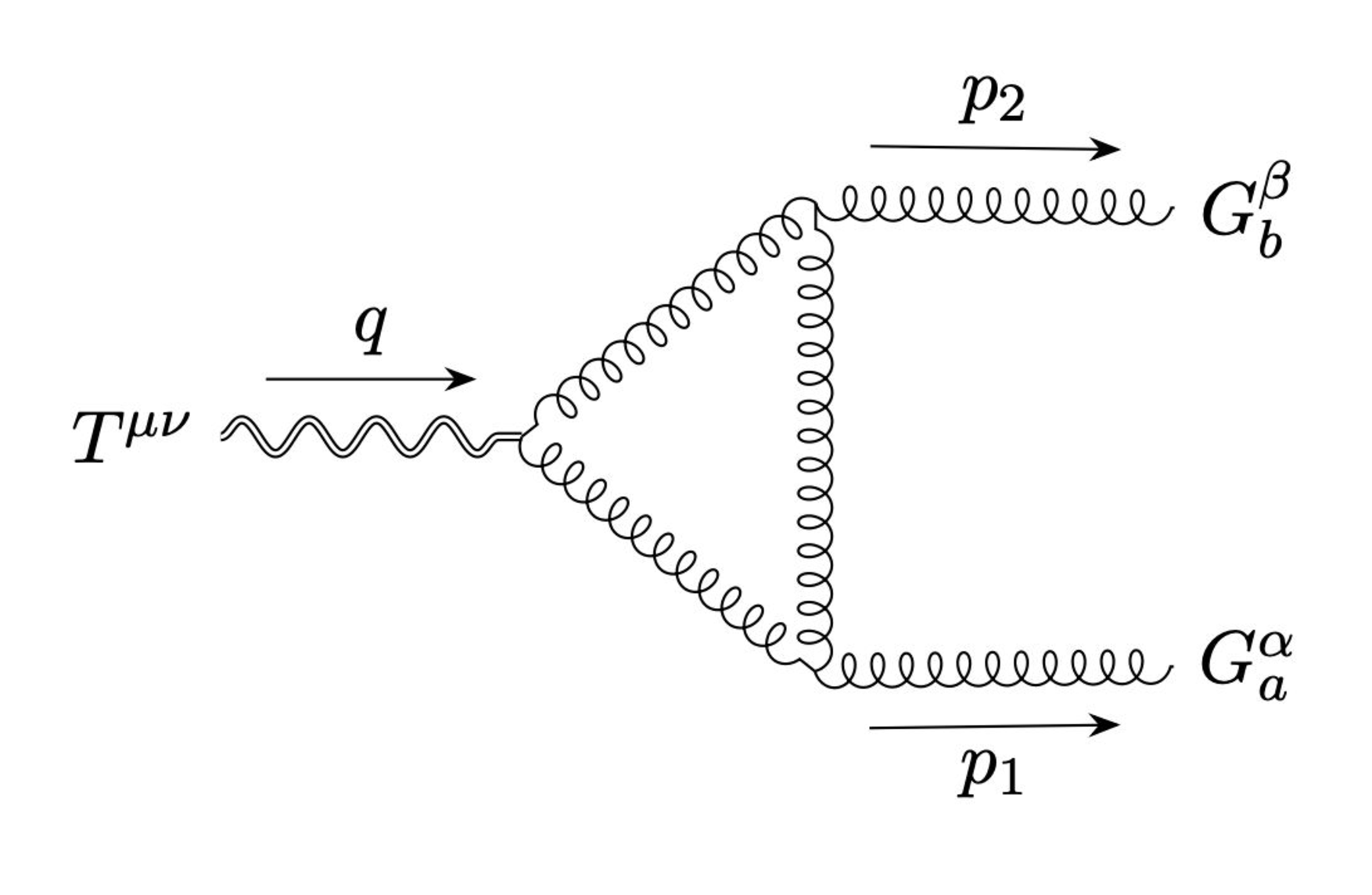}}
\rotatebox{-90}{\includegraphics[scale=0.1]{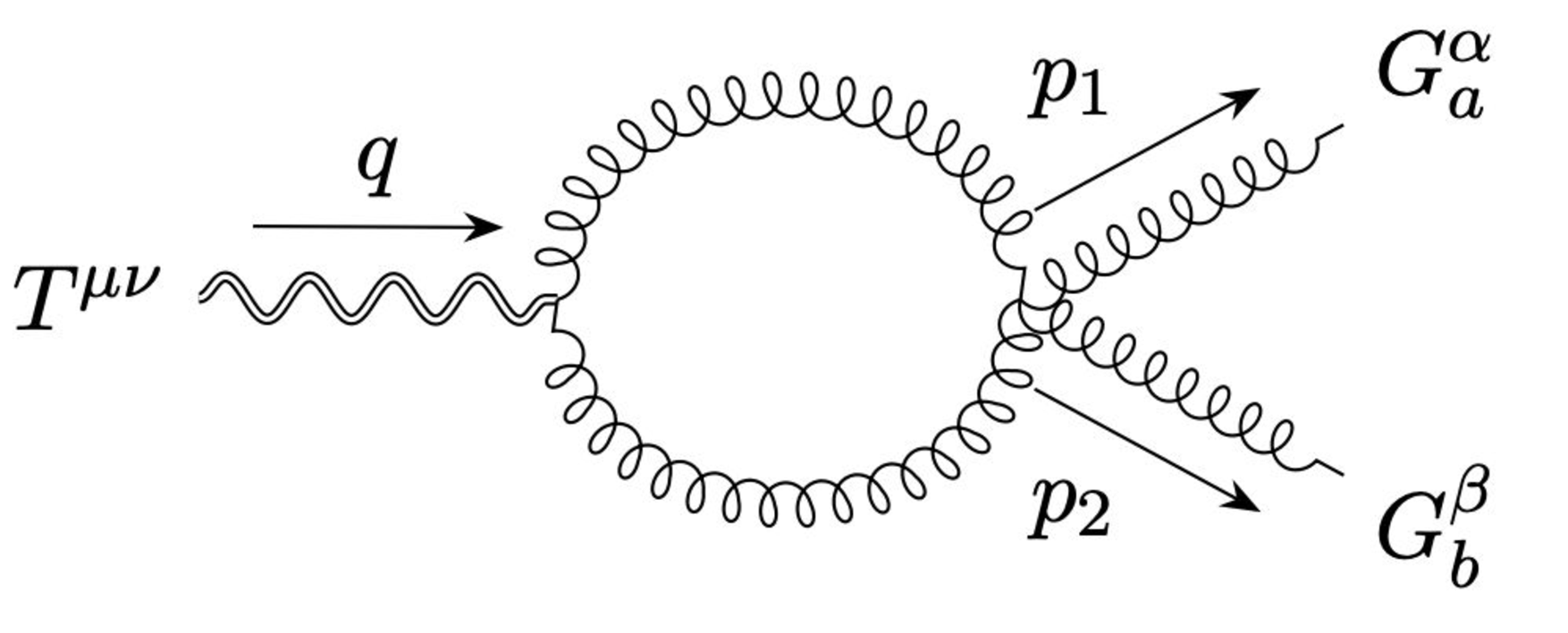}}
\caption{Typical contributon to the TJJ vertex at one loop, present in the quark and gluon sectors.}
\end{center}
\end{figure}
\section{Conclusions}
The conformal anomaly interaction is described through a longitudinal, transverse, and trace sector decomposition of the relevant vertices, with a pole emerging in the trace channel. This parameterization of hard scattering processes, as presented, can be seamlessly extended to the hadronic level, specifically for protons and pions, offering a potential phenomenological framework for analysing the role of this anomaly in the GFFs of pion and proton. We have shown that conformal anomalies are intrinsically tied to the presence of effective dilaton degrees of freedom in hard scattering processes. In the perturbative framework, these anomalies manifest as the emergence of a dilaton pole during hard scattering. This phenomenon is accompanied by a sum rule, satisfied by the anomaly form factor, which we have checked at the one-loop level in perturbation theory, in analogy to other anomaly form factors for chiral interactions. Details of this specific analysis can be found in 
\cite{Coriano:2024qbr}.
\vspace{0.5cm}
\centerline{\bf Acknowledgements}
This work is partially funded by the European Union, Next Generation EU, PNRR project "National Centre for HPC, Big Data and Quantum Computing", project code CN00000013; by INFN, inziativa specifica {\em QG-sky} and by the grant PRIN 2022BP52A MUR "The Holographic Universe for all Lambdas" Lecce-Naples.


\end{document}